# An ELIXIR scoping review on domain-specific evaluation metrics for synthetic data in life sciences

Styliani-Christina Fragkouli [1,2], Somya Iqbal [3], Lisa Crossman [4,5], Barbara Gravel [6,7,8], Nagat Masued [9], Mark Onders [10], Devesh Haseja [11], Alex Stikkelman [12], Alfonso Valencia [9,13], Tom Lenaerts [6,7,8], Fotis Psomopoulos [2], Pilib Ó Broin [11], Núria Queralt-Rosinach [12], Davide Cirillo[9]

1 Department of Biology, National and Kapodistrian University of Athens, Athens, 15772, Greece
2 Institute of Applied Biosciences, Centre for Research & Technology Hellas, Thessaloniki, 57001, Greece
3 University of Edinburgh, Usher Institute, Nine BioQuarter, Edinburgh EH16 4UX. UK
4 SequenceAnalysis.co.uk, Norwich Research Park, Norwich, NR4 7UG, UK
5 School of Biological Sciences, University of East Anglia, Norwich Research Park, Norwich, NR4 7TJ, UK
6 Interuniversity Institute of Bioinformatics in Brussels, Université Libre de Bruxelles-Vrije Universiteit Brussel, 1050 Brussels, Belgium
7 Machine Learning Group, Université Libre de Bruxelles, 1050 Brussels, Belgium.
8 Artificial Intelligence Laboratory, Vrije Universiteit Brussels, 1050 Brussels, Belgium.
9 Barcelona Supercomputing Center (BSC), Plaça Eusebi Güell, 1-3, 08034, Barcelona, Spain
10 University of Notre Dame, Notre Dame, IN 46556, United States
11 School of Mathematical and Statistical Sciences, College of Science and Engineering, University of Galway, Galway, Ireland
12 Department of Human Genetics, Leiden University Medical Center, Leiden, 2333, The Netherlands
13 ICREA, Pg. Lluís Companys 23, 08010, Barcelona, Spain

# corresponding author: davide.cirillo@bsc.es

# Abstract

Synthetic data has emerged as a powerful resource in life sciences, offering solutions for data scarcity, privacy protection and accessibility constraints. By creating artificial datasets that mirror the characteristics of real data, allows researchers to develop and validate computational methods in controlled environments. Despite its promise, the adoption of synthetic data in Life Sciences hinges on rigorous evaluation metrics designed to assess their fidelity and reliability.

To explore the current landscape of synthetic data evaluation metrics in several Life Sciences domains, the ELIXIR Machine Learning Focus Group performed a systematic review of the scientific literature following the PRISMA guidelines. Six critical domains were examined to identify current practices for assessing synthetic data. Findings reveal that, while generation methods are rapidly evolving, systematic evaluation is often overlooked, limiting researchers' ability to compare, validate, and trust synthetic datasets across different domains.

This systematic review underscores the urgent need for robust, standardized evaluation approaches that not only bolster confidence in synthetic data but also guide its effective and responsible implementation. By laying the groundwork for establishing domain-specific yet interoperable standards, this scoping review paves the way for future initiatives aimed at enhancing the role of synthetic data in scientific discovery, clinical practice and beyond.

# Introduction

Synthetic data (SD) is trending across fields to address the lack of real data (RD), privacy concerns and accessibility limitations while enabling robust machine learning (ML) and artificial intelligence (AI) applications and benchmarking studies [1]. By generating artificial datasets, SD allows researchers to develop and validate computational methods in controlled environments by facilitating the evaluation of statistical models and predictive algorithms. However, as technologies continue to advance, the volume and complexity of biological and clinical data have grown significantly and while this expansion presents new opportunities for discovery, it also introduces challenges in ensuring data availability, quality and privacy protection. A promising solution lies with SD by enabling the creation of high-fidelity datasets that bypass privacy restrictions while preserving essential statistical properties [2].

Beyond practical considerations, SD is valuable when strict data governance regulations regulate access to RD. Legal frameworks like the General Data Protection Regulation (GDPR) and the Health Insurance Portability and Accountability Act (HIPAA) impose stringent requirements on sharing and using sensitive data in healthcare research [3]. SD offers a potential viable solution by sharing synthetic versions of patient records that maintain statistical properties of RD while mitigating privacy risk. This facilitates inter-institutional collaborations by reducing bureaucratic hurdles and ensuring compliance with ethical and legal standards [4].

The effective integration of SD into scientific and clinical workflows depends on its ability to accurately mirror the structure of RD. Evaluating SD quality remains a major challenge, as its utility is tied to assessment methods that verify its structural and inferential fidelity. Evaluation needs vary by application: some prioritize enhancing ML/AI model performance, while others aim to uncover novel patterns for hypothesis generation. Consequently, there is no one-size-fits-all evaluation method, assessment strategies must align with the goals and context of each use case [5].

To address these challenges, robust evaluation recommendations and best practices are essential for assessing SD across diverse applications. Much like ongoing initiatives in areas like cross-domain FAIR data assessment and the Critical Assessment of Structure Prediction, the development of guidelines for SD could progressively lead to a more structured and coherent framework for evaluating prediction methods across domains. Such guidelines will enhance the trustworthiness of SD, ensuring its reliability in biomedical research and clinical decision-making [6]. Furthermore, as SD continues to evolve, its successful integration into healthcare will require more rigorous evaluation recommendations and clear regulatory guidelines as well as ongoing adaptation to uphold its reliability and ethical implementation in medical research and practice.

The ELIXIR ML Focus Group is part of ELIXIR, the European infrastructure for life sciences, which brings together over 250 research organizations across 24 countries. In this work, a collaborative effort was led by the ELIXIR ML focus group, specifically the Task Force on SD, in order to establish an evaluation set of guidelines for SD in life sciences within Europe. A scoping review based on the Preferred Reporting Items for Systematic reviews and Meta-Analyses (PRISMA) standards was carried out by experts in the Task Force across different domains adjacent to life sciences in order to identify the current evaluation methods and metrics that are utilized to assess SD (individual flow diagrams for each domain are available in Supplementary Figures S1–S6). The key domains selected for the review are genomics, transcriptomics, proteomics, phenomics, imaging, and electronic health records (EHRs), ensuring a thorough and comprehensive investigation of the topic. **Figure 1** displays a word cloud visualization of the metrics identified across the six domains, illustrating the diversity of the metrics in a visual format (with all metrics visualized collectively in Supplementary Figure S7).

This scoping review maps current methodologies and highlights gaps in evaluation strategies to support the systematic assessment of SD quality, reliability, and utility in life sciences research. While existing studies provide in-depth surveys of SD quality metrics in biomedicine, they tend to focus on specific aspects, like privacy and utility [7], benchmarking general-purpose implementations [8], or particular healthcare domains collectively [9]. To our knowledge, this is the first comprehensive review that systematically examines evaluation metrics for SD across multiple application domains in the Life Sciences, addressing each domain and in detail.

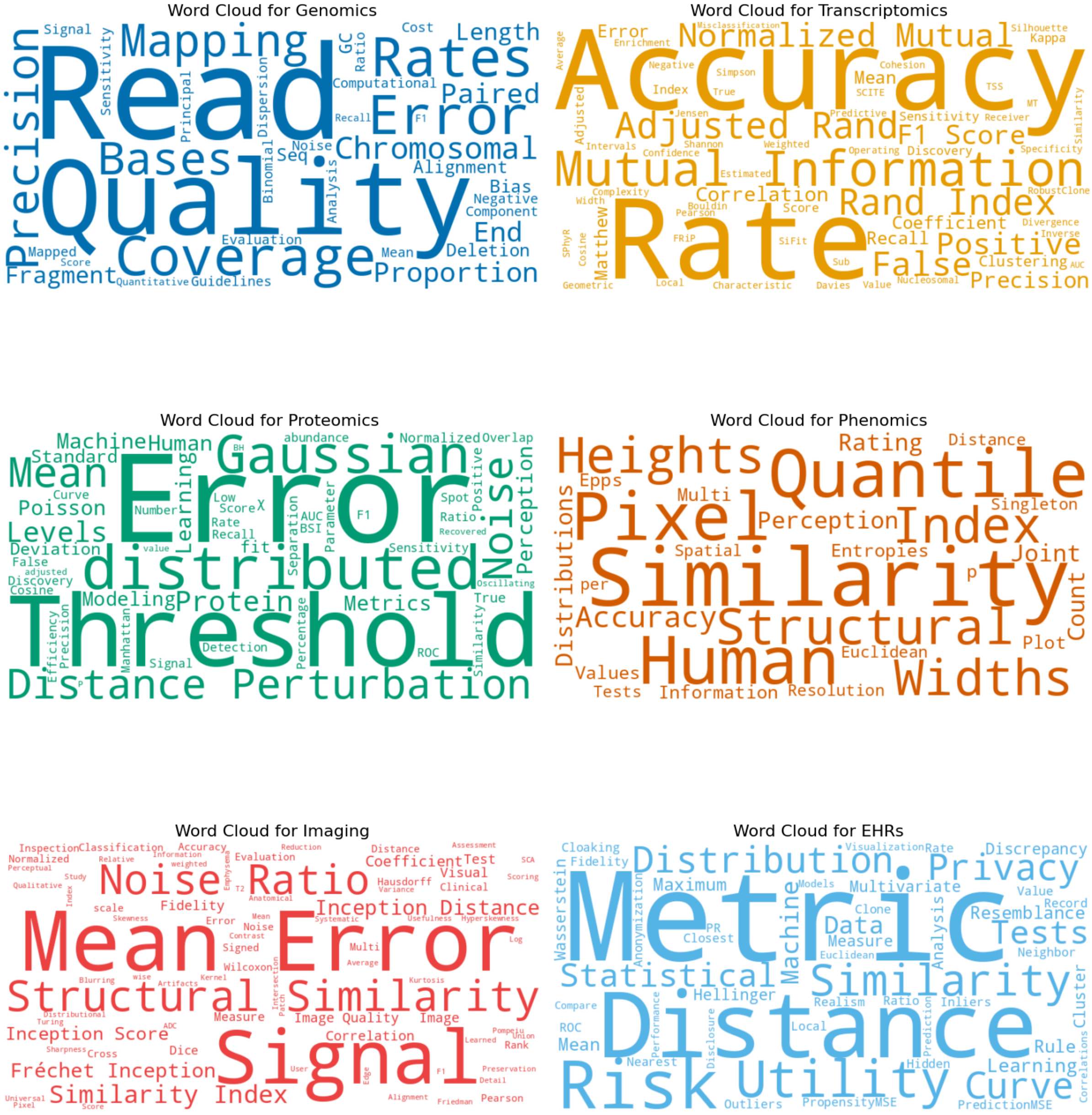

**Figure 1.** Word cloud visualizations of the metrics identified across genomics, transcriptomics, proteomics, phenomics, imaging and EHRs.

# Methods

## Inclusion Criteria

Following the PRISMA [10] guidelines, we devise a systematic approach to identifying relevant literature on SD across various domains, ensuring the inclusion of high-quality, peer-reviewed, and openly accessible research. To identify relevant studies, we conducted a comprehensive literature search using inclusion criteria as shown in the PRISMA diagram for each domain (Supplementary Figures S1–S6 and Supplementary Material).

The selection of studies for this scoping review was guided by predefined inclusion criteria to ensure a comprehensive yet focused assessment of evaluation metrics for SD. Literature searches were conducted across multiple databases, including PubMed, SCOPUS and Google Scholar, while Web of Science (WoS) and DOAJ.org as optional sources as their specialized focus might not align with all research requirements. To maintain relevance to current research trends only studies published within the last ten years were included.
All selected publications were in English and accessible as Open Access articles to facilitate transparency and reproducibility. Furthermore, only peer-reviewed journal publications were considered to ensure the scientific rigor and credibility of the included studies. Given the large volume of unspecific results typically obtained by Google Scholar using queries, those search results were ranked by relevance [11], with the top 20 articles reviewed for potential inclusion. This approach ensures that the review captures high-quality, methodologically sound literature that contributes meaningfully to the evaluation of SD (see **Table 1**).

**Table 1**: The inclusion criteria used for our comprehensive literature search in order to identify relevant studies.

| **Criteria** | **Details** |
|---|---|
| Databases | PubMed, SCOPUS, Google Scholar, Web of Science (WoS) (optional), DOAJ.org (optional) |
| Timeframe | Studies published in the last 10 years |
| Language | English |
| Accessibility | Open Access |
| Peer-reviewed | Articles from peer-reviewed journals |
| Ranking | In the case of Google Scholar, top 20 results ranked by relevance |

## Search Queries

A different set of queries were formulated, tailored to two key aspects; firstly to accurately reflect the specific domain of the evaluation and secondly to align with the search engine used in each instance. All queries incorporated the terms “synthetic” and “evaluation”, followed by domain-specific keywords to ensure relevance for each domain. The primary domains deliberately chosen for exploration include genomics, transcriptomics, proteomics, phenomics, imaging and EHRs in order to ensure a comprehensive investigation of the topic. For further details please see the Supplementary Material file.

## Identification of relevant publications

The PRISMA 2020 flow diagram, depicted for each domain search in Supplementary Figures S1–S6, provides a structured representation of the study selection process in systematic reviews that involve searches from databases and registers. It consists of three main stages: identification, screening and inclusion. During the identification stage, records are collected from multiple sources including databases and registers. Before the screening process begins, duplicate records are removed and additional exclusions may occur (as described later in the limitations section) due to automation tools or other filtering criteria.
During the screening stage, the remaining records undergo an initial review to determine their relevance. At this step, reports are sought for retrieval and any that cannot be accessed are documented. The retrieved reports are then assessed for eligibility and exclusions are made based on specific reasons, which are clearly recorded (i.e. not relevant for evaluating SD).
Finally, in the inclusion stage, the studies that meet all criteria are incorporated into the systematic review and the total number of included reports is documented. This method ensures transparency and reproducibility by visually summarizing the number of records processed at each stage and detailing exclusions thus enhancing clarity. All domain related PRISMA flowcharts can be found in the Supplementary Material (Figures S1-S6).

## Limitations

During the search process, several observations and challenges were identified. The term "synthetic" often refers to synthetic biology rather than SD generation, leading to unrelated findings. Additionally, many studies mention "synthetic data" in abstracts without detailing evaluation metrics or generation tools. The number of relevant papers varies significantly by domain, with fewer studies found in complex fields like synthetic phenomics. Furthermore, additional care was necessary to consider studies focused on synthetic peptides and related data generation from those as opposed to computationally generated SD in proteomics. Although the majority of the literature search and curation were conducted in the first half of 2024, we continuously reviewed additional papers in each domain and found no additional relevant metrics beyond those that we had already surveyed. This strengthens the authors' confidence that this review accurately reflects the current state of the art.

# Results

## Synthetic Genomics

### Domain overview

Synthetic genomics involves using computational tools to design, analyze and simulate synthetic genomes. It aims to create accurate representations of genetic sequences by leveraging bioinformatics and computational biology along with wet-lab expertise. Some key concepts include designing genomes computationally, modeling the functionality of the synthesized genomes, utilizing this type of data for the development and the benchmarking of bioinformatics tools. Designing optimized genomic sequences for specific functions is driving advancements in vaccine development [12] as well as synthetic and evolutionary biology [13]. However, challenges like data limitations, biosecurity risks, and model interpretability persist, highlighting the need for solutions to ensure safety and ethical compliance.

## Review outcomes

After the examination of 769 articles, 6 publications were identified appropriate for our effort. These selected articles fall under two main categories, reviewing genomics simulating tools and suggesting guidelines for the evaluation [14]. Due to the broad use of the term 'synthetic,' many irrelevant publications (either outside of the inclusion criteria or due to technical aspects that were unsuitable for this review) were initially identified, requiring us to refine the queries with more appropriate search terms. Furthermore, the term “synthetic DNA” typically refers to data that is generated in wet labs rather than in-silico.

## Evaluation metrics

The reported metrics can be divided into two groups; those that can be found and utilized in a standard bioinformatics pipeline and/or analysis and those tailored to specific use cases.

- **Bioinformatics related metrics** [15–17]: proportion of reads on each chromosome, coverage, type of reads, paired-end fragment length, alignment quality, rates of sequencing error, substitution, insertion, and deletion rates, quality scores for reads, gc-coverage bias, computational cost, mapping sensitivity of bases in reads, mapping precision of mapped bases in reads, Phred score.
- **Use case related metrics** [18]: PCA to evaluate qualitatively by visualization and supervised discriminators for quantitative evaluation.
- **Statistics based metrics** [19]: mean of significant variables in the case/control, the dispersion parameter of the negative binomial distribution, and the signal to noise ratio, sample size, number of variables, mean of insignificant variables.

# Synthetic Transcriptomics

## Domain overview

Synthetic transcriptomics is a key research area within the field of SD generation for biological studies, enabling the creation and analysis of artificial RNA-Seq data that mirrors the characteristics of RD [20]. Such SD can aid in standard downstream transcriptomics analyses like the identification of differentially expressed genes and isoforms and pathway enrichment analysis [6,21], and more advanced applications, like the identification of clinically-relevant biomarkers. Synthetic transcriptomic data is often used in the context of oversampling approaches to correct imbalance between different sample groups within a dataset – this can help to compensate for bias due to underrepresentation of smaller sample groups, lead to increased statistical power and reduce the effect of outliers [22]. A key factor contributing to the increasing use of synthetic transcriptomic data relates to the fact that the process of RNA isolation, library preparation and sequencing to obtain RD is relatively time, labour, and cost intensive, making the alternative approach of generating realistic SD attractive in terms of both efficiency and conservation of resources [23]. This is particularly true for the adoption of synthetic transcriptomic data for the training of ML algorithms where adequate sample size, and restricted access to RD can be a concern.

## Review outcomes

The search criteria for ‘synthetic transcriptomic’ or ‘synthetic RNA’ resulted in 424 articles (139 in scopus, 23 pubmed, 16 web of science, 246 google scholar). After removing duplicate records (n=14), additional filtering was carried out to exclude articles not meeting other criteria (articles in other languages, preprints or conference proceedings (n=120), articles found to be more than 10 years old (n=3), articles belonging to a different domain (n=33). Ultimately, 84 articles remained for curation and the resulting evaluation metrics from these are summarised below.

A key aspect of SD evaluation in this domain is the distinction between intrinsic and extrinsic metrics. Extrinsic evaluation refers to performance-based metrics that assess how well a model behaves when trained on SD compared to RD. Common extrinsic metrics include precision, recall, F1-score, AUC-ROC and other standard measures used in ML/AI. These metrics are particularly relevant when assessing the usability of SD in real-world applications. On the other hand, intrinsic evaluation focuses on the inherent quality of SD itself, independent of any downstream task. These metrics assess characteristics like distribution similarity, perplexity (e.g., in the case of SD for language models) and various statistical measures that capture how closely SD resembles RD. One approach to intrinsic evaluation involves confusion matrices, which can provide insight into how different SD generation methods relate to one another, helping to quantify similarities or systematic biases. By combining both intrinsic and extrinsic evaluation methods, a more comprehensive understanding of SD quality can be achieved.

## Evaluation metrics

Many different evaluation parameters were identified during the review. Here, we have attempted to segregate them into different categories according to their underlying properties.

- **Quantitative Metrics:** These metrics are used for comparison of quantitative parameters between the SD and RD. These can be subclassified as follows:.
  - Performance Metrics: These are the standard statistical metrics that are generally used to test the performance of the data generation models. Examples include Sensitivity, F1 Score, Precision, AUC-ROC score, AUC-PRC score, Confusion matrix.
  - Classification metrics: These metrics are used to benchmark the classification accuracy of the model. They consist of parameters like Matthews Correlation Coefficient (MCC) and Cohen’s Kappa.
  - Clustering metrics: The clustering metrics assess the quality of clusters formed from SD. Examples are Adjusted Rand Index (ARI) and Normalised Mutual Information (NMI).
  - Ranking metrics: They are used to compare the rank of the data generated by the model in order of relevance. Examples are Spearman’s rho and Kendall’s tau.
  - Regression metrics: These metrics evaluate the predictive ability of a given model. Examples are Huber’s Loss and Error rate.
- **Biologically-informed/Use case Metrics:** These metrics assess the difference between the biological properties of the SD and RD. Examples include: read coverage, read count, read depth and related read metrics like RPKM and FPKM, overlap in annotation/classification of single cells, fold change in gene/isoform expression and clustering accuracy.
- **Qualitative Metrics:** These metrics focus on the comparison of the aspects of data generating algorithms. Examples include: Speed, Complexity of datasets, Privacy, Diversity.

# Synthetic Proteomics

## Domain overview

Proteomics falls under the biological study of proteins and is embedded in the wider Omics field. The breakdown of proteomics can be split into: quantitative proteomics, functional proteomics, structural proteomics, protein-protein interactions, qualitative methods with mass spectrometry imaging (MSI), proximity extension assays, and aptamer based platforms amongst others [24]. The instrumentation and technical methodologies also span multiple formats, with emerging single cell proteomics making a foray [25] . Comprehensive overview of the listed methodologies and their nuanced analytical approaches are described in [26,27] for further reading. Data generated from Quantitative Mass Spectrometry (qMS) proteomics, includes spectral peaks, abundance values, intensity signals in MS, amounts of protein in samples, and many other measurement outputs depending on sub-field and instrumentation, which are of relevance when discussing SD in this domain.

## Review outcomes

We identified 129 initial articles and 68 after duplicate removals, which went through an abstract screening phase, before the full text review (n=57), and then data extraction (n=15). The initial searches used combined queries per database (see Supplementary Material), Pubmed n= 52, Scopus n=35, Google Scholar n=24 (top 20 by relevance plus second query), Web of Science (n= 16), DOAJ (n=6), total 129 papers. The reviewed papers did not have SD generation as the core focus, but rather SD was used as part of wider research questions, and evaluation metrics were extracted from multiple sections to infer any measures taken to evaluate the data.

## Evaluation metrics

The reported metrics span the broad themes of quantitative, biological and qualitative sub categories, an example of sub category types would be statistical measures for quantitative metrics like FDR during identification and quantification stages of MS spectra, for qualitative metrics, this would include expert inspections, overlays, benchmarking, whilst biological would include similarity measures against biological benchmarks, closeness to true outputs from protein/peptide measures and or synthetic peptide standards.
The papers included reported evaluations of the SD produced, and further details around how they assessed these data in relation to overall study goals.

Types of SD generated or used in the included papers:
1. Synthetic mass spectra - spectral data, spectra, isobaric peptides [28–34]
2. Synthetic 2-dimensional gel electrophoresis (2DGE) spots - images [35,36]

3. Synthetic expression data, synthetic features of proteomic data for designated models (high dimensional formats, signal intensities, perturbations, cut-offs) [37–40]
4. RPPA, P100 assays, and global chromatin profiling (GCP) [38]
5. Human Protein Atlas (HPA), large-scale immunohistochemistry (IHC) [41]
6. Networks (biological with proteins) [41,42]

The main metrics and most represented area of proteomics from the papers was from MS based proteomics research.

**Quantitative metrics (including statistical and performance):**

- Signal-to-Noise Ratio (SNR) - image quality metric expressed in dB (image based SD), Spot Efficiency - evaluating spot detection performance (image based SD). FDR (true spots versus false - image based SD), Subtraction index with background pixels, Sensitivity, true spots detected divided by total true spots. (image based SD), Low-abundance Protein Detection (LPD), percentage of low-abundance spots detected. Other ML related metrics.
- Gaussian-distributed Perturbation, Stepwise perturbations, simulating calibration uncertainty in peak fitting, Modelling of acquisition windows, noise and collision based classifications (MS), Standard Deviation of Gaussian fit - measuring precision in peak intensity fitting, Normalized Separation Parameter ($\chi$ = dt / HWHM) - for peak separation analysis.
- True Positive Distance Threshold - peak is a true positive if within 0.3% m/z of true peak, Cosine Similarity Threshold - isotope pattern matching threshold, thresholding variations in expression, Manhattan Distance threshold - peptide matching threshold set at 10 units.
- Noise Levels / Perturbation Levels - evaluated as factors influencing AUC, sensitivity, and specificity.
- Protein expression levels and effect sizes (modulation to simulate RD)
- AUC (area under the curve), ROC (receiver operating characteristic curve) - model efficacy
- Poisson-distributed Error Modeling - simulating counting errors with variance proportional to sqrt(N).
- Feature matrix - Alignment, peak binning, peak detection, Randomised noise for predictor features, exponential noise
- F1 Score - harmonic mean of (1 - FDR) and Sensitivity.
- ROC Curve - showing TPR (True Positive Rate) vs. FPR (False Positive Rate).
- MSPE (Mean Square Prediction Error) - for dose-response model comparison.
- Data distributions: Binomial distribution for feature clusters, Assortativity distributions (networks), Random sampling of distributions, Markov chain Monte Carlo (MCMC)
- False Discovery Rate (FDR) - false spots divided by total spots detected.
- BSI - pre-processing performance measure.
- Recall (Recovery Rate) - fraction of recaptured initial seed nodes or disease proteins.
- Precision - true positives divided by (true positives + false positives).
- MAE (Mean Absolute Error) - used in network noise reduction performance.
- BH-adjusted P-value - used to assess model classification significance (cutoff $< 0.05$).
- Canonical coefficients related to SD weights

- Balancing of datasets during evaluation of models, replicate consistency, batch effects

**Qualitative Metrics:**

- Subjective (Perceptual) Evaluation - visual assessment of image fidelity to real 2DGE images.
- Visual data distribution graphs and counts between RD and SD
- Generating replicates - similarity
- Manual inspection of peptides and lengths (amino acids)
- network connectivity and node size to simulate RD
- Use of ground truth labels

**Biological and domain specific Metrics:**

- SNR - capturing heterogeneity in expression data
- Protein abundance values/ranges, peptide lengths, isotope pattern fidelity
- Pooling for minimising isobaric peptide signals
- representation of disease associated proteins (% known, subsets), module overlaps
- Ground truth with biological sets.
- Similarity (true molecular ions)

**Comparison metrics:**

- Overlap Percentage - comparing model predictions with seed perturbation scenarios.
- Number of Oscillating Proteins Recovered - metric from MOSAIC modeling.
- Comparing model performance on RD and SD (characteristics and features specific to study context) - PSM, identifications, counts correct number of classifications).
- Graphed data comparisons, distributions, counts and relative differences under varying conditions (perturbations, noise, thresholds).

Metrics like ROC, AUC, MSPE, and F1 were the most suitable metrics for assessing SD for model based validations, whilst noise and perturbations were used to mimic real time variations and heterogeneity, the LogFC thresholds thereby maintained relevancy to downstream analytics and classification. Ground truth recovery was identified as a way to ensure robustness in noisy network based paradigms, with grid searches allowing for many simulations improving reproducibility, whereas those employing manual inspection or subjective evaluation provide expert interaction with quantitative metrics. Metrics like pooling and identifying isobaric peptide signals as well as peptide length are considered contextually relevant items related to MS instrumentation.

Of the metric types recorded, in the context of proteomic data the comparison metrics would be deemed bespoke in that the model applications, which were designed to be validated themselves were used to show comparative metrics between RD and SD, with the main goal of highlighting the efficacy of the model as opposed to the SD generation. However, those which carefully considered instrumentation parameters and key features from RD could still provide a richer view of the SD quality metrics. The metrics used for algorithmic performance should ideally be comparable between SD and RD.

Some papers focused on SD tailored to specific model evaluations like linear or correlation models without aiming to replicate RD. These SD were often designed for targeted validation tasks rather than broad mimicry of RD. However, reporting on how well the SD captured the intended characteristics was often insufficient. In cases where SD generation methods were adopted from prior work, descriptions of the process and evaluation were frequently minimal.

Although the use of existing methods is common, clearly reporting the conditions and assumptions remains important. Furthermore, the review did not apply snowballing techniques, making it difficult to retrieve details when SD generation was only referenced indirectly.

Research papers using synthetic peptides for quantification required nuanced reviewer skills, the generation of synthetic peptides has existing and established quality control metrics, however papers which used synthetic spectra from synthetic peptide quantification for model evaluation purposes and provided evaluations either on how they then used MS to quantify alongside the RD or how they assessed these data in their own right were included for those metrics.

Many of the excluded papers had little to unclear reporting of evaluation metrics which required further inference. No papers in the search outputs focused solely on producing SD for proteomics as the overarching goal of the research, however there was a critical evaluation paper on the use of artificial spectra data which is already successfully used in MS peptide identifications and considering these SD for training algorithms in de novo peptide identification. Whilst the paper offers further reading on the important additions of noise from MS measurements and a need for bespoke SD for de novo methodologies, providing useful insights for the domain, the content focussed on evaluating the artificial training data and compared model capabilities via peptide recall and was not focussed on independently evaluating the SD data itself. Extracting SD quality metrics could not be easily identified outside of the prediction algorithm space since this was a critical evaluation of SD applicability in de novo peptide identifications [43]. Further expert commentary on the ML development in proteomics with relevance to SD and SD generation in the field can be found elsewhere [44,45].

The range of papers included in the review represent a wide scope of the types of data and contexts available to proteomics and the extracted metrics for SD quality can provide an insightful index for future standardisation.

# Synthetic Phenomics

## Domain Overview

Phenomics is a relatively new discipline of Biology that has been applied in several fields, mainly in crop sciences [46,47] . Phenomics is the systematic study of the complete set of expressed phenotypes and the structure of such a set in relation to the genetic factors and environmental perturbations that caused their expression [48]. It combines concepts and methods from various disciplines to understand the complex interactions between genotype, phenotype, and environmental factors that influence organismal development, function, and evolution [49,50]. While traditional approaches to phenomics rely heavily on experimental measurements [51], synthetic phenomics offers a complementary strategy that harnesses the power of computation to create simulated phenotypes.
Synthetic phenomics is a subfield within the broader scope of phenomics, focusing on the creation and utilization of SD to investigate complex traits. By leveraging state-of-the-art computational models and simulation techniques, we can now generate vast quantities of SD that closely resemble phenotypes from RD. This innovative approach has the potential to significantly enhance our understanding of complex traits, like those involved in cancer, neurological disorders, and metabolic diseases [52]. Overall, phenomics has the potential to transform our understanding of complex diseases and traits, paving the way for the development of personalized therapies and targeted interventions.

## Review outcomes

From the initial large number of publications (i.e., a total of 6920 identified from the queries on PubMed (n=2), Web of Science (n=21), Google Scholar (n=6890), and ELIXIR curation registry (n=7)) only 22 were finally selected for the review after the screening process. Before screening, we applied an extra filter to reduce the untreatable number of records obtained from Google Scholar to the top 20 ranked per relevance. This resulted in 50 records to be screened. We excluded 5 synthetic phenomics datasets records from the ELIXIR curated registry for not providing a peer-reviewed supporting publication. From the 45 retrieved reports, we excluded 3 publications for not being open access, and 20 for being false positives, .i.e., they did neither generate nor use SD. We observed that 8 out of these 20 false positives were due to the ‘synthetic biology’ topic of the publication. Noteworthy, of the final 22 publications we reviewed, a significant number, the 91% (20/22), fall into the Plant Sciences domain, and the majority of publications, the 32% (7/22), were issued in 2021.

## Evaluation metrics

Only 3 publications from the total reviewed publications, i.e., 14%, mentioned the use of evaluation metrics to assess the quality of the SD used or/and generated. The metrics identified were both quantitative and qualitative and overall dependent on the modality of the generated dataset:

1. **Quantitative metrics** [53]: widths and heights in pixels of real vs synthetic maize tassels images; structural similarity (SSIM) index computation to indicate the similarity between two images, real and synthetic; usability of SD by accuracy score of a classifier model.
2. **Perceptual qualitative metrics** [53]: human perception annotation about real or synthetic image; human rating on the similarity of a pair real-synthetic image.
3. **Count distributions similarity** [54]: distributions of the count table elements.
4. **Distributions of information measures** [54]: joint entropies and any information theoretic measure which is a function of the entropies such as multi-information Ω.
5. **Statistical testing of distribution equivalence** [54]: quantile-quantile plot of p-values from Epps-Singleton tests comparing synthetic versus expected distributions.
6. **Realism of the simulated dataset** [55]: Euclidean distance.
7. **Image spatial resolution** [55]: spatial resolution per pixel.

# Synthetic Imaging

## Domain Overview

Synthetic imaging is a rapidly evolving field in medical imaging, leveraging advanced computational techniques to generate realistic medical images. These synthetic images are pivotal in various applications, including training ML/AI models, enhancing diagnostic accuracy, and facilitating the development of new imaging technologies. Synthetic imaging presents numerous advantages and diverse applications in the medical field [56]. One key benefit is data augmentation, where synthetic images enrich existing datasets by providing a wider variety of samples, effectively addressing the limitations of RD. This approach also enhances privacy protection, as SD eliminates the need to use sensitive patient information, while simultaneously reducing the costs and effort associated with annotating real medical images. Furthermore, synthetic imaging serves as a valuable tool for training and educating healthcare professionals, offering a risk-free environment that safeguards patient confidentiality. Beyond education, it opens up new avenues for research, like modality translation and the development of AI-driven diagnostic tools.

## Review outcomes

A comprehensive review of synthetic imaging literature from the past decade was conducted, which identified and screened a large number of publications. The initial search outcomes were: PubMed identified 13 articles, with 10 fitting the criteria after applying the last 10 years filter; SCOPUS identified 21 articles, with 18 fitting the criteria after filtering; and Google Scholar identified 307 articles, with 296 fitting the criteria after applying the last 10 years filter.

After examining 341 articles, 45 articles were finally identified as an appropriate fit for our effort. These selected articles fall under two main categories: those reviewing tools for simulating synthetic images and those suggesting guidelines for evaluation. The initial search yielded a large number of publications with only a few that were relevant, mostly due to the broad usage of the term “synthetic”, prompting the need for more precise search terms.

## Evaluation metrics

Evaluating the quality and performance of synthetic images involves both quantitative and qualitative metrics.

**Quantitative metrics** as identified in the curated publications [57–93] include:

- **Peak Signal-to-Noise Ratio (PSNR)**: Measures the ratio between the maximum possible power of a signal and the power of corrupting noise, indicating image quality.
- **Structural Similarity Index (SSIM)**: Assesses the similarity between two images, focusing on luminance, contrast, and structure.
- **Mean Squared Error (MSE)**: Calculates the average squared differences between the generated and reference images.

- **Mean Absolute Error (MAE)**: Measures the average absolute differences between the pixel values of the synthetic and real images.
- **Fréchet Inception Distance (FID)**: Evaluates the distance between the distributions of real and synthetic images in a feature space, providing an overall measure of image quality.

**Use case related metrics**: These metrics are tailored for specific applications:

- **Visual Inspection:** Expert radiologists visually assess image quality, anatomical detail, and clinical utility.
- **Radiologist Evaluation:** Radiologists use rating scales to assess image sharpness, contrast, and artifact presence.

**Statistics based metrics**: These metrics are derived from statistical analyses:

- **Wilcoxon Signed-Rank Test:** A non-parametric test to compare paired samples, like synthetic and real images.
- **Friedman Test:** A non-parametric test to detect differences in treatments across multiple test attempts.
- **Principal Component Analysis (PCA):** Used for qualitative data visualization.
- **Supervised Discriminators:** Employed for quantitative evaluation, measuring how well synthetic images can be distinguished from real ones.

# Synthetic EHRs

## Domain overview

EHRs serve as an extensive, lifelong repository of patient health information, encompassing diverse data types like medical history, medications, allergies, lab results and more. These records, often dispersed across multiple platforms and locations, provide a detailed, chronological overview of a patient's medical journey and facilitate data sharing across healthcare settings. However, the fragmented nature of EHR systems complicates access to patient records, creating challenges for clinical practitioners, researchers, and developers [94].

## Review outcomes

After curating 83 articles, 18 of them were finally selected. Regarding the generation methods the main sources mainly involve types of GANs, probabilistic models, classification models, weighted bayesian association rule mining, Gaussian multivariate, bayesian networks and sequential tree synthesis. As indicated by [94] and further supported by other works [3,95–97], different evaluation metrics are often utilised, usually based on the type of generation method that was initially implemented to produce SD, which makes the direct comparison of the results quite challenging. On the other hand, the work of [98] distinguishes the metrics used based on data type focusing on tabular data and time series.

## Evaluation metrics

A plethora of metrics are utilized for the evaluation of EHR SD. It is important to clarify that this literature is primarily targeting the structured part of the EHRs, and therefore the textual form is not taken into consideration (nor any AI-drive aspects). As such, the SD in this case are tabular data and sometimes longitudinal data. The majority of the publications that were selected indicated three main categories of evaluation metrics regarding the resemblance of the SD, the utility and finally the preservation of privacy. Furthermore, many studies note the distinction of the latter category into further information disclosure related subcategories. The first subcategory includes those regarding the identity disclosure which occurs when an attacker can accurately determine that a specific individual is part of the training dataset. This risk arises when one or more SD points closely resemble a RD sample used in generating the synthetic dataset, potentially revealing its inclusion in the original confidential data. The second subcategory includes those that deal with the attribute disclosure which refers to the risk that an attacker can accurately predict the original values of sensitive attributes for an individual in the confidential dataset. This occurs when the attacker uses known variables from the RD, combined with patterns in the SD, to infer the sensitive values. The likelihood of such disclosure depends on factors like the number of known variables, the size of the synthetic dataset, and the configuration of the attack model.

The metrics that were identified after carefully curating the papers that were included in the manuscript are categorized and presented below:

- **Resemblance evaluation**: variable distributions, frequency of data features, dimensional probability or probability distributions, Consultation with clinical experts (qualitatively), Data similarity; Synthetic to Synthetic (STS), Real to Synthetic (RTS) and Real to Real (RTR) , Pairwise Pearson correlation coefficient, most common values, Statistical similarity (mean, std, miss rate), Visualisation techniques (PCA, histograms and correlation matrices), Nearest neighbor adversarial accuracy (AA) and resemblance loss, average trends, Train an ML classifier to label data as real or synthetic, distribution of attributes, Kolmogorov–Smirnov, Welsch t-tests, Student t-test, Chi Square tests, Dimension-wise distribution similarity (Dimension-wise Probability, Dimension-wise Average, Kolmogorov–Smirnov (K-S) test, Support Coverage, Kullback–Leibler Divergence (KLD)) [99–102].

- **Utility evaluation**: Augment data for ML model training, Use SD in ML models, Kullback-Leibler (KL) divergence, pairwise correlation difference (PCD), log-cluster, support coverage, cross-classification, Multivariate Hellinger distance, Wasserstein Distance, Cluster analysis measure, Propensity and prediction MSE, Dimension-wise distribution, Column-wise correlation, Latent cluster analysis, Clinical knowledge violation, Medical concept abundance, Feature selection [103,104].

- **Privacy evaluation**: Identity disclosure and attribute disclosure, Distance to the closest record (DCR), Membership attack, Max-RTS similarity, Formulation of DP, Privacy loss, Maximum mean discrepancy (MMD), JS divergence (JSD) and Wasserstein distance, Differential privacy cost, Distance to the optimal point , Attribute inference risk, Membership inference risk, Meaningful identity disclosure risk, Nearest neighbor adversarial accuracy risk [94,101,103–105].

- **Extra metrics**: ML performance metrics (Sensitivity, Specificity, Positive Predictive Value/Precision, Negative Predictive Value, AUC score, Precision–Recall AUC score), Rule similarity (for specific data type generation) [105–107].

# Discussion

Our review of evaluation metrics for SD across six distinct Life Sciences domains (genomics, transcriptomics, proteomics, phenomics, imaging, and EHRs) highlights both the diversity and domain-specific nature of the metrics commonly used in the field. While SD is playing an increasingly important role in scientific research, evaluating its quality and applicability remains a significant challenge, particularly when comparing across these diverse domains. Although SD is tailored to specific applications within each domain, the lack of standards makes it difficult to assess and compare SD quality even within the same application domain. This challenge becomes even more pronounced with the growing interest in multimodal applications of SD, where multiple data types from different domains are artificially generated and merged. This trend highlights the opportunity to develop new evaluation metrics tailored to these complex use cases, which will require a separate review as applications begin to develop further.

In total, after the systematic screening of over 8,000 initial records retrieved from multiple databases 188 publications were finally curated across all SD domains. The refinement process involved multiple layers of filtering, including duplicate removal, exclusion of preprints and non-peer-reviewed material, non-English texts, articles older than ten years and records that misused or ambiguously applied the term "synthetic." A total of 156 metrics were identified (**Figure 2**), of which 142 are unique to a single domain and 14 are shared across multiple domains. These shared metrics are illustrated in **Figure 3** and explored in greater detail in Supplementary Figure S8.

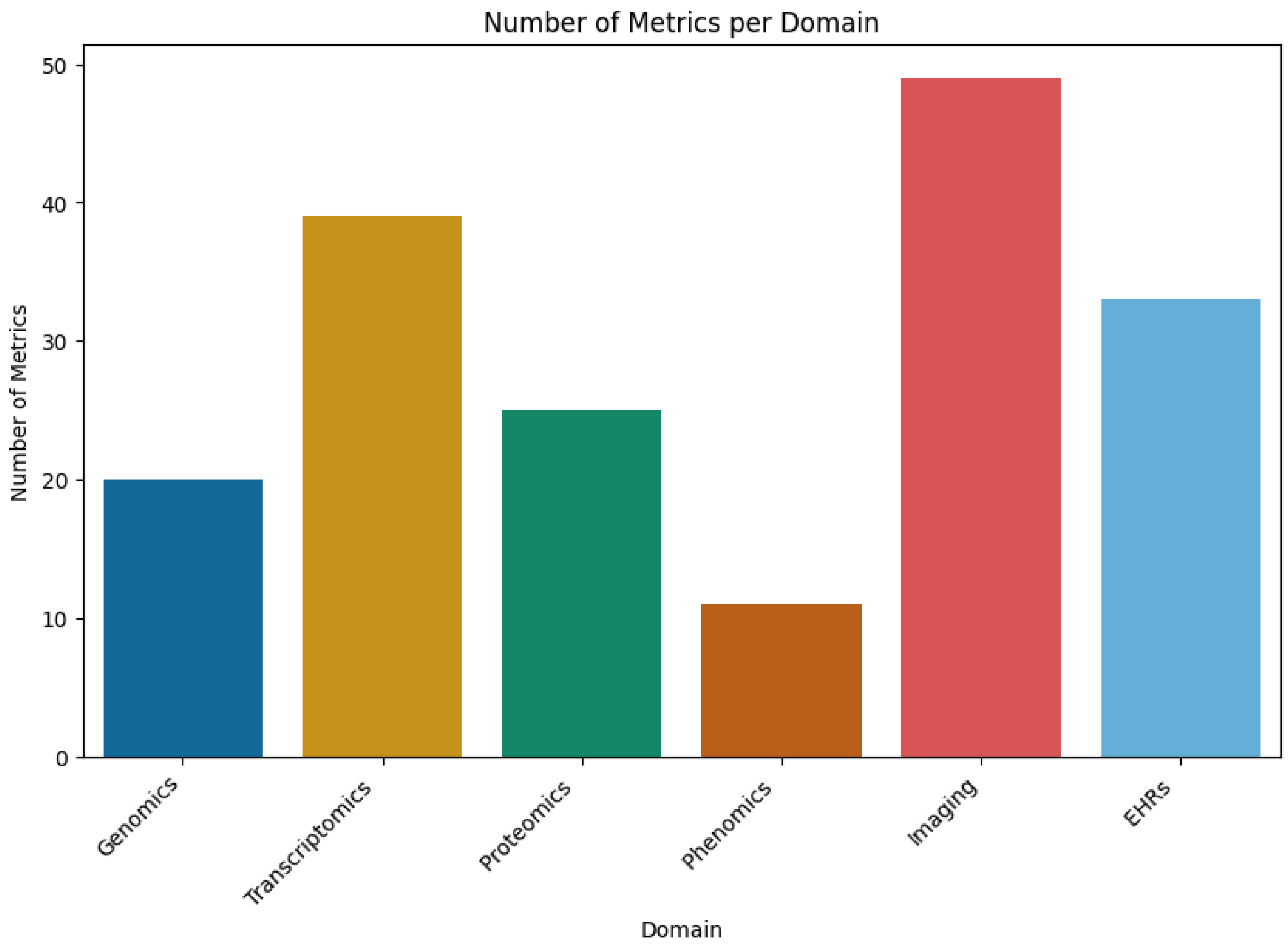

**Figure 2.** Barplot showing the number of SD evaluation metrics that were retrieved for each domain.

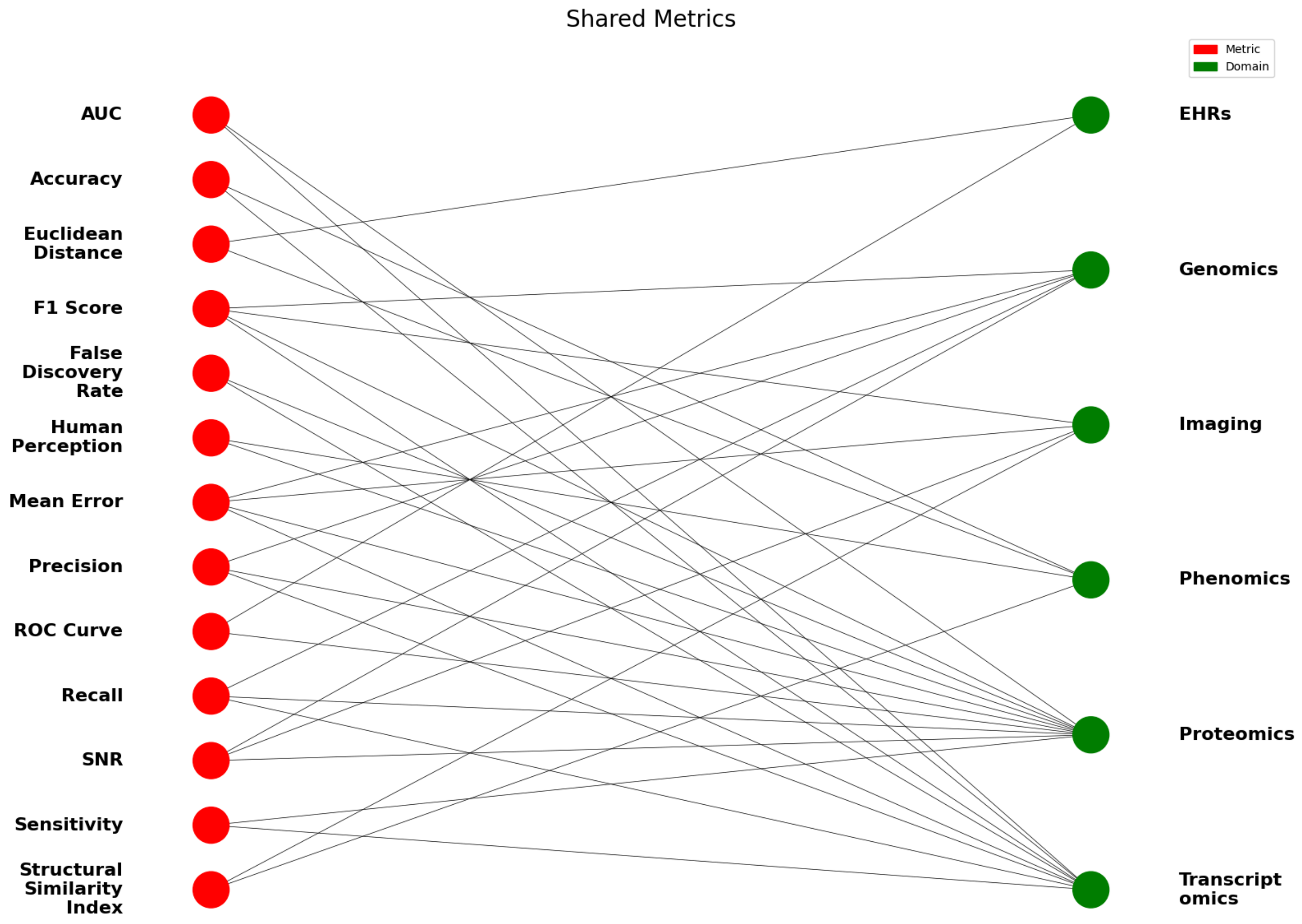

**Figure 3.** Bipartite graph displaying evaluation metrics (red nodes) that are shared across multiple domains (green nodes).

In the domain of **genomics** the review found that the term "synthetic" has often a different meaning, with many studies referring to wet-lab synthesized DNA rather than computationally generated sequences. This issue underscores the need for clearer terminology and domain-specific criteria for evaluating SD in genomics. Regarding the curation effort, the outlined metrics can be put into three categories; the bioinformatics-related metrics, use case-related metrics and finally statistics-based metrics. Similarly, in **transcriptomics**, SD plays a crucial role in gene expression analysis and ML/AI applications. The primary focus for the application of SD in this domain include data augmentation and balancing underrepresented classes to increase statistical power for the downstream analyses. Extrinsic metrics assess model performance when trained on SD versus RD, using measures like precision, recall, F1-score and AUC-ROC, which are crucial for real-world applicability. Intrinsic metrics, on the other hand, evaluate the inherent quality of SD, focusing on distribution similarity, perplexity and statistical comparisons. While extrinsic metrics ensure that SD is suitable for practical applications, intrinsic metrics provide fundamental insights into data fidelity, consistency and variability. By combining both approaches, a more comprehensive and robust SD assessment can be achieved, contributing to more reliable benchmarking frameworks.

In **proteomics** SD was employed in many of the reviewed papers, whilst the generation process was described, the data itself and evaluation of its use was not often described with clear metrics. However, from the papers reviewed, a number of metrics were identified which could point towards some agreement to develop best practices for evaluating the SD within the sub categories of proteomics research methodologies. The criteria were context dependent, in the case of 2DGE for instance which utilises images, compared to MS spectra based SD, relying on analysis tool based outputs primarily and comparison of model performance. Some of the more distinct metrics were ROC, AUC, peak accuracy/matching, sensitivity, specificity and ML/image fidelity metrics for 2DGE.

Clear ontologies and frameworks for defining SD in scientific domains, as well as the tasks performed using SD and related evaluation metrics are much needed to guide effective domain specific applications. In the case of proteomics, with multiple iterative steps from sample curation to data interpretation, the workflows have many stages dependent on the methodologies employed to identify proteins. For qMS, there are many points in the workflow to consider the efficacy of SD, like parameters used in MS instrumentation which vary from data driven acquisition (DDA), data independent acquisition (DIA) approaches, labelling versus non labelled approaches, processing softwares, and varying spectra features like peaks, noise, fragmentation patterns, as for downstream outputs expression data would be required to represent features like coverage, FDR, and the dimensionality of the data typical of outputs.

The review excluded direct evaluation of synthetic peptide production, as this area follows established standards distinct from in silico SD generation. The role of SD in proteomics is to augment for the purpose of predictive modelling and not to replace RD. In essence, appropriately generated SD with evaluation and quality metric protocols in place can offer exploratory research spaces to improve and simulate refined questions of interest, prior to employing experimental examinations.

In **phenomics**, the reviewing process found that while SD offer a promising approach for studying genotype-phenotype interactions, the primary focus remains on data generation rather than systematic assessment, indicating a gap that needs to be addressed to ensure the reliability of SD for this domain. Furthermore, the dominance of plant-based studies suggests a potential bias in SD research, requiring further exploration in human-related phenomics applications. The metrics here can be divided into various categories, including qualitative, perceptual qualitative, distribution similarity, information measure, statistical testing of distribution equivalence, realism of SD and image spatial resolution.

Medical **Imaging** is one of the more mature fields in terms of SD applications, as synthetic medical images are widely used for training AI models and augmenting RD. However, despite the significant number of studies reviewed, most publications focused on the development of synthetic imaging tools rather than comprehensive evaluations of their reliability. Metrics like structural similarity and image fidelity were commonly reported, but there was a lack of consistency in benchmarking methodologies. The metrics that were identified as relevant in this domain can be split into quantitative, use case-related and statistics-based metrics.

Lastly, in the domain of **EHRs**, SD is used to generate privacy-preserving patient records for research and AI model training. The review found a wide variety of generation methods, including GANs, probabilistic models and Bayesian networks, each employing different evaluation metrics. The lack of uniformity in assessment strategies makes it difficult to establish clear guidelines for evaluating synthetic EHR data, further emphasizing the need for a standardized approach. At this point it should be noted that both Medical Imaging and EHR SD applications are particularly relevant in a clinical context, and the respective need for more rigorous and standardized evaluation ways to assess the clinical utility of the SD in these fields.

A key challenge across domains is the inconsistency in evaluation criteria, which often depends on the SD generation method. Some studies emphasize statistical similarity to RD, while others focus on downstream utility, like improving ML/AI performance. Additional approaches assess data integrity, reproducibility, or real-world applicability. This diversity highlights the need for domain-specific yet practical evaluation guidelines that ensure consistent, rigorous, and meaningful assessments tailored to each domain's unique needs.

As part of a broader, ongoing effort to enhance the visibility and accessibility of synthetic data resources in the life sciences, we are working towards contributing the ELIXIR SD collection to the BioStudies registry [108]. This planned contribution aligns with the development of a Registry of Life Sciences Synthetic Datasets, which aims to standardize access to SD across domains. It further complements the initiatives of the ELIXIR ML Focus Group, particularly the Task Force on SD, which is actively developing evaluation guidelines for assessing the quality, utility, and applicability of SD in life science research across Europe.

Furthermore, as members of ELIXIR Europe [109], we advocate for interoperability in SD evaluation metrics that align with existing standards and related initiatives. This approach is consistent with the software best practices defined under the ELIXIR-STEERS project [110] and the software quality indicators developed within the European Open Science Cloud (EOSC) ecosystem [111]. For example, the EVERSE project [112] aims to establish a framework for research software excellence by integrating community curation, quality assessment, and best practices, which will be encapsulated in the Research Software Quality toolkit (RSQkit) as a comprehensive knowledge base for high-quality software development across disciplines. The interoperability of SD evaluation metrics could be achieved by developing dedicated ontologies and semantic strategies to represent the evaluation metrics to allow their integration in SD generation benchmarking platforms, such those currently developed in European projects like SYNTHIA [113] and SYNTHEMA [114].

Despite advancements in medical AI research, the clinical adoption of emerging AI solutions remains challenging due to limited trust and ethical concerns. Several recent initiatives are contributing to the development of robust best practices and guidelines for the evaluation of SD quality and its applications in the Life Sciences. Similarly to current efforts in areas like the assessment of FAIR data across domains, or the Critical Assessment of Structure Prediction, these best practices and guidelines for SD could evolve into a more structured framework for the evaluation and assessment of prediction methods across domains. The FUTURE-AI consortium [115] has defined international guidelines for trustworthy healthcare AI addressing technical, clinical, socio-ethical and legal aspects. These recommendations ,

built on six principles (fairness, universality, traceability, usability, robustness, and explainability), spans the entire AI lifecycle and ensures alignment with real-world needs and ethics, promoting multi-stakeholder collaboration and continuous risk assessment. Additionally, the AHEAD project [116] seeks to create a global community of experts from diverse fields, including biomedicine, ethics, AI development, sociology, and law, to develop guidelines for evaluating AI-based systems in healthcare. It focuses not only on technical performance but also on human and societal factors, addressing ethical concerns, privacy implications, fairness, transparency, and the broader societal impact.
We believe that by integrating both technical and societal evaluation approaches and aligning with initiatives like FUTURE-AI and AHEAD, more robust and comprehensive recommendations for assessing SD can be established. Such an integrated approach would not only promote responsible AI innovation but also ensure that the deployment of SD across various domains is both reliable and aligned with broader societal values.

# Conclusions

This ELIXIR scoping review underscores the critical need for standardized evaluation metrics to assess the quality, reliability, and applicability of SD across diverse scientific domains.

Key findings from this review highlight that in most domains, research efforts are primarily focused on developing SD generation methods rather than systematically evaluating their effectiveness. This imbalance suggests that future research should prioritize the establishment of robust evaluation methodologies to ensure the credibility of SD in scientific and clinical applications.

To fully leverage the potential of SD, future efforts should focus on defining clear guidelines for assessing synthetic datasets, with a particular emphasis on establishing shared evaluation standards. This will not only enhance the comparability of different SD generation techniques but also improve confidence in the use of SD across multiple research disciplines. The findings of this review serve as a foundation for future initiatives aimed at standardizing SD evaluation, ultimately ensuring its responsible and effective integration into scientific research and healthcare applications.

# Acknowledgments

The authors would like to acknowledge The ELIXIR Machine Learning Focus Group for helpful discussions. Also, the authors would like to express their gratitude to Magnus Palmblad, Rahuman Sheriff, Sara Morsy, Ruben Branco, Tim Beck, Salvador Capella-Gutiérrez, and Sucheta Ghosh for their valuable support and insightful advice.

# Authors contribution

DC and NQR conceived and designed the review. SCF, SI, BG, NR, DH, AT, LC, and NQR conducted the literature search, analyzed and interpreted the findings, and contributed to manuscript writing. AV, TL, FP, and POB supervised the work. All authors reviewed and approved the final manuscript.

# Funding

This work was supported by ELIXIR, the research infrastructure for life-science data. The Centre for Research & Technology Hellas (CERTH), Barcelona Supercomputing Center (BSC), and Leiden University Medical Center (LUMC) receive support from SYNTHIA. SYNTHIA (Synthetic Data Generation framework for integrated validation of use cases and AI healthcare applications) is supported by the Innovative Health Initiative Joint Undertaking (IHI JU) under grant agreement No 101172872. The JU receives support from the European Union's Horizon Europe research and innovation programme, COCIR, EFPIA, Europa Bío, MedTech Europe, Vaccines Europe and DNV. The UK consortium partner, The National Institute for Health and Care Excellence (NICE) is supported by UKRI Grant 10132181. Funded by the European Union, the private members, and those contributing partners of the IHI JU. Views and opinions expressed are however those of the author(s) only and do not necessarily reflect those of the aforementioned parties. Neither of the aforementioned parties can be held responsible for them.

# Authors short description

Styliani-Christina Fragkouli is a PhD candidate at the National and Kapodistrian University of Athens and a research associate at INAB|CERTH, focusing on synthetic data generation, somatic variant calling and machine learning applications in multiple domains.

Somya Iqbal is a PhD researcher at the University of Edinburgh, specialising in Mass Spectrometry based proteomics with a focus on data-driven (ML and network science) analysis workflows for biomarker discovery in health and disease.

Lisa Crossman is Director of SequenceAnalysis.co.uk, with over a decade of experience in microbial genomics consulting and a PhD in Microbiology. Her expertise includes metagenomics, RNA-seq, bioinformatics and translational research in life sciences.

Barbara Gravel holds a PhD in Bioinformatics from the Vrije Universiteit Brussel and the Université Libre de Bruxelles and focuses on oligogenic inheritance, variant prioritization and machine learning methods to identify complex genetic contributions to disease.

Nagat Razeg is a Research Engineer in the Machine Learning for Biomedical Research Unit at the Barcelona Supercomputing Center (BSC), contributing to projects at the intersection of AI and life sciences.

Mark Onders is a Senior at the University of Notre Dame majoring in Applied and Computational Mathematics and Statistics with a focus on Scientific Computing. His interests lie in software engineering, data science, and machine learning.

Devesh Haseja is a research assistant at the University of Galway, contributing to research in precision medicine and functional genomics by applying statistical and machine learning methods to integrate genetic and clinical data.

Alex Stikkelman is an active member of SV Nucleus and is involved in research on synthetic genomics data, focusing on innovative approaches to genomic simulation and analysis.

Alfonso Valencia is Director of the Spanish National Bioinformatics Institute. His group develops advanced machine learning tools for the integration and analysis of large-scale genomic and clinical data, contributing to major international consortia such as ENCODE, ICGC, BLUEPRINT, and RD-Connect.

Tom Lenaerts is a Professor at the University Libre de Bruxelles and the Vrije Universiteit Brussel. He is co-heading the Machine Learning Group, specializing in interdisciplinary AI research with applications in collective intelligence, computational biology, and personalized medicine, including oligogenic disease modeling.

Fotis Psomopoulos is a researcher at INAB|CERTH. His research interests include bioinformatics and machine learning, primarily working on the design and implementation of novel algorithms for knowledge extraction from large datasets in life sciences.

Pilib Ó Broin is an Assistant Professor in Bioinformatics at the University of Galway, leading research in precision medicine and functional genomics, integrating genetic and clinical data using statistical and machine learning methods.

Núria Queralt-Rosinach is a biomedical informatics researcher at Leiden University Medical Center developing semantic and AI-based methods to explore disease mechanisms and the genotype-environment-phenotype interplay for improved therapeutic understanding.

Davide Cirillo is Head of the Machine Learning for Biomedical Research Unit at BSC. His work focuses on computational methods for precision medicine, combining machine learning, network science, and AI ethics.

# Key Points

- Synthetic data generation methods in life sciences (spanning in genomics, transcriptomics, proteomics, phenomics, imaging and EHRs) have advanced rapidly, but systematic evaluation of their effectiveness remains limited, with most studies focusing on generation rather than rigorous assessment .

- There is a critical need for standardized, shared evaluation metrics to assess the quality, reliability and applicability of synthetic datasets across diverse scientific domains, ensuring their credibility in both research and clinical contexts .

- The review categorizes existing metrics into resemblance, utility and privacy evaluations, yet highlights a lack of consistency in benchmarking methodologies that impedes cross-domain comparability .

- Integrating technical performance with socio-ethical considerations will be essential for developing trustworthy guidelines throughout the synthetic data lifecycle .

- Establishing clear ontologies and contributing evaluation frameworks to centralized registries will facilitate responsible, effective integration of synthetic data into life-science research and applications .